\documentclass[reprint,amsmath,amssymb,aps,prl]{revtex4-2}
\usepackage{graphicx}
\usepackage{dcolumn}
\usepackage{bm}
\usepackage{xcolor}
\begin{document}

\title{Giant asymmetric second-harmonic generation in bianisotropic metasurfaces based on bound states in the continuum}
\author{Ehsan Mobini$^{*1}$, Rasoul Alaee{$^{*1,2}$}, Robert W. Boyd{$^{1}$, and Ksenia Dolgaleva$^{1}$}}
\address{
{$^{1}$}Department of Physics, University of Ottawa, 25 Templeton, Ottawa, Ontario, K1N 6N5, Canada\\
$^{2}$Karlsruhe Institute of Technology, Institute of Theoretical Solid State Physics, Wolfgang-Gaede-Str. 1,
D-76131 Karlsruhe, Germany\\
$*$ $\rm{Corresponding\,\, authors: rasoul.alaee@gmail.com,\,\,\, eh.mobini@gmail.com}$
}

\begin{abstract}
Bianisotropy is a powerful concept enabling asymmetric optical response, including asymmetric reflection, absorption, optical forces, light trapping, and lasing. The physical origin of these asymmetric effects can be understood from magnetoelectric coupling and asymmetrical field enhancement. Here, we theoretically propose highly asymmetric second-harmonic generation (SHG) in bianisotropic AlGaAs metasurfaces. We show that around \textit{four} orders of magnitude second-harmonic power difference for the forward and backward illuminations can be obtained by altering
geometrical parameters that coincide with quasi-bound states in the continuum. Our study paves the way towards directional generation of higher-order waves and can be potentially useful for nonlinear holograms.
\end{abstract}

\maketitle
\section{Introduction} 
Among planar photonic devices, metasurfaces have experienced a revolutionary path of development over the past decade, enabled by state-of-art fabrication methods and rich physics~\cite{Chen2016}. The ability to control the amplitude, phase, and polarization of reflected, transmitted, and absorbed light was pioneered by plasmonic metasurfaces~\cite{meinzer2014plasmonic,huang2020planar}. However, the inherent Ohmic loss of plasmonic metasurfaces is a limiting characteristic for the linear and nonlinear functionalities. An inspiring idea to overcome this challenge emerged by employing periodic arrangements of lossless high-index dielectric meta-atoms~\cite{kuznetsov2016optically,kruk2017functional,rybin2017high,baranov2017all}. Due to the high confinement of light supported by the electric and magnetic Mie resonances, dielectric metasurfaces found its unique place in nonlinear optics~\cite{camacho:2016,makarov:2017,kruk2017nonlinear,Kruk:2019,Xu:2019,Smirnova:2020}. In addition, the optically thin nature of the dielectric entities releases the device from phase matching. 
In this context, the physics of dielectric metasurfaces can be explained in terms of coupled-multipolar interactions~\cite{evlyukhin2012demonstration,Smirnova:16,liu2017multipolar,alaee2018electromagnetic,Alaee2019}. 


A migrated quantum mechanical concept to the optics scope, known as optical bound states in the continuum (BICs), has been widely used as an efficient approach to enhance the confinement of light inside the meta-atoms~\cite{hsu2016bound,koshelev2020engineering}.
Optical bound states in the continuum are optical modes that are unable to couple with the allowed continuum of radiative modes in the surrounding medium~\cite{hsu2016bound}. Although ideally BICs necessitate an infinite Q-factor, one can observe a finite Q-factor by breaking the symmetry of the meta-atoms, so-called quasi-BICs~\cite{koshelev2018asymmetric}. To date, thanks to this concept, enhancement of the conversion efficiency of the second-~\cite{carletti2018giant,liu2019high,anthur2020continuous,kang2021efficient}, third-~\cite{koshelev2019nonlinear,zhou2020resonant,gandolfi2021near}, and higher- order~\cite{carletti2019high,zograf2020high} harmonics in dielectric metasurfaces have been demonstrated theoretically and experimentally. The physical origin of these large nonlinear conversion efficiencies can be understood by the field enhancement~\cite{volkovskaya2020multipolar}.
Controlling the direction and phase of the generated higher-order waves, which is crucial for nonlinear holography~\cite{huang2018metasurface,gao2018nonlinear} and nonlinear wavefront shaping~\cite{wang2018nonlinear,gigli2021tensorial}, is another significant relationship between dielectric metasurfaces and nonlinear optics.

In terms of controlling the wave direction, bianisotropy is a powerful concept enabling an asymmetric behavior of the interacting light with the scatterer based on the magnetoelectric coupling~\cite{lindell1994electromagnetic,asadchy2018bianisotropic}. A strong bianisotropic response can be achieved by breaking the inversion
symmetry~\cite{alaee2015all,odit2016experimental}.  Up to now, bianisotropic metasurfaces have demonstrated asymmetric linear functionalities such as asymmetric reflection~\cite{pfeiffer2014bianisotropic,ra2014tailoring}, asymmetric optical forces~\cite{alaee2015all}, self-isolated Raman lasing~\cite{dixon2021self}, light trapping~\cite{evlyukhin2020bianisotropy}, asymmetric absorption~\cite{Alaee:15,yazdi2015bianisotropic,Alaee:2017Review}, and nonreciprocity~\cite{wang2020nonreciprocity,cheng2021superscattering}, among other feats~\cite{lindell1994electromagnetic,asadchy2018bianisotropic}. However, the nonlinear potentials of bianisotropic metasurfaces based on bound states in the continuum have remained unexplored.

\begin{figure*}
\centering
\includegraphics[width=0.75\textwidth]{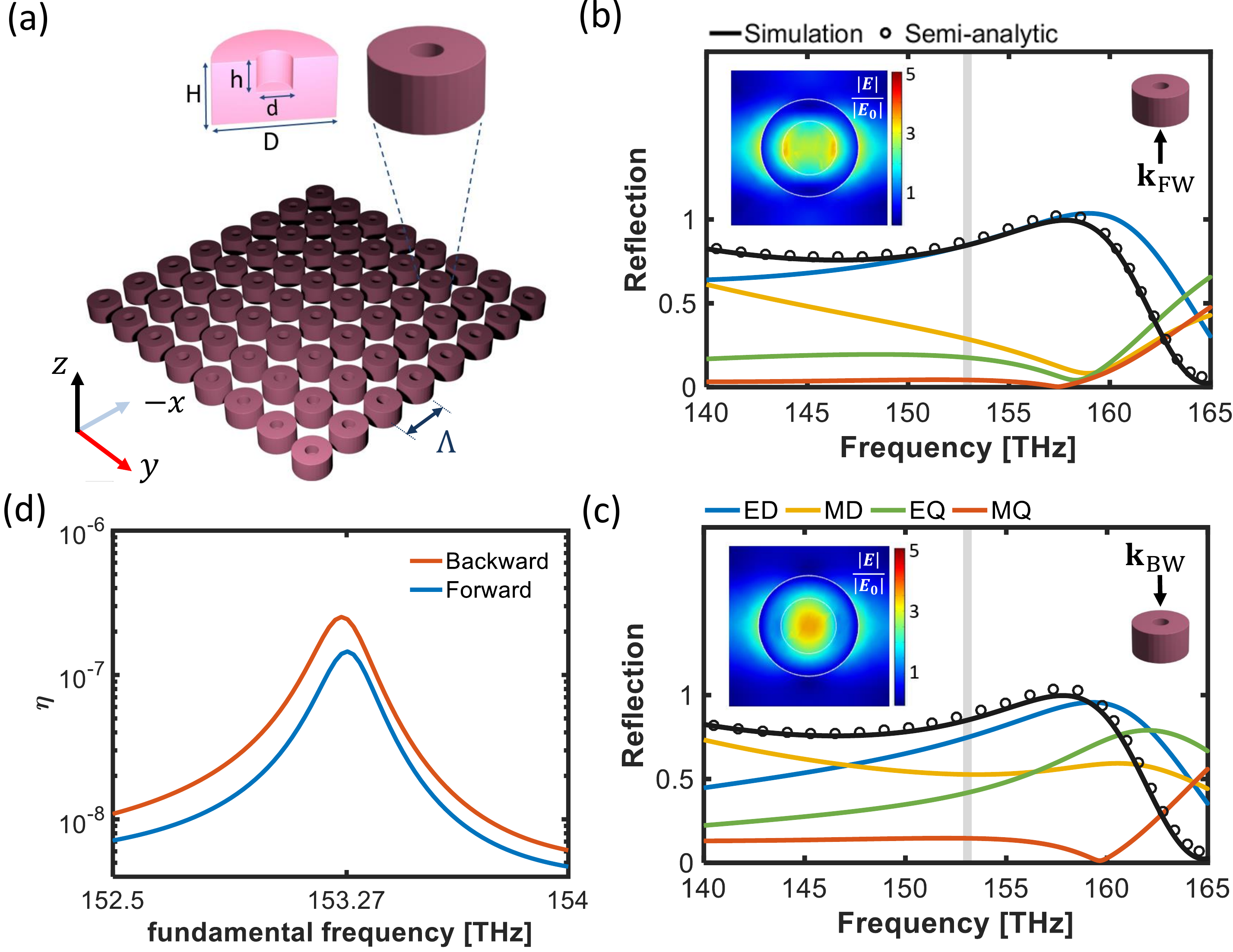}
\caption{Linear and nonlinear responses of the bianisotropic metasurface. (a) Schematic of the bianisotropic metasurface with a coaxial cylindrical hole and its cross-sectional view with the geometrical dimensions. The metasurface is illuminated by a linearly polarized light with wavevector $\mathbf{k}_{\rm FW/BW} = \pm k_0\mathbf{e}_z$, where $+$ and $-$ indicate forward and backward illuminations, respectively. (b-c) The reflection spectrum and multipolar contributions for the forward and backward illumination directions, respectively. The semi-analytic reflection is obtained using Eq.~(\ref{Eq_ref}).
The insets show the normalized electric field $\left|\mathbf{E}\right|/\left|\mathbf{E}_{0}\right|$ in the middle of the meta-atom ($xy$-plane), where $\left|\mathbf{E}_{0}\right|$ is the amplitude of the incident field. The grey shaded bar
shows the frequency at which the SH conversion efficiency is investigated. ED (MD) and EQ (MQ) indicate contributions of electric (magnetic) dipole and quadrupole moments, respectively. (d) The second harmonic conversion efficiencies $\,\rm \eta=P_{\rm SHG}/P_{\rm pump}$ for the forward and backward illumination directions. The geometrical
parameters of the bianisotropic metasurface are $ d=420\,\rm{nm}$,\,$ D=760\,\rm{nm}$, $ h=360\,\rm{nm}$,
$ H=720\,\rm{nm}$ and $ \varLambda=1260\,\rm{nm}$, where $\varLambda$ is the periodicity in $x$ and $y$ directions.}  \label{fig1}
\end{figure*}

\begin{figure*}
\centering
\includegraphics[width=0.75\textwidth]{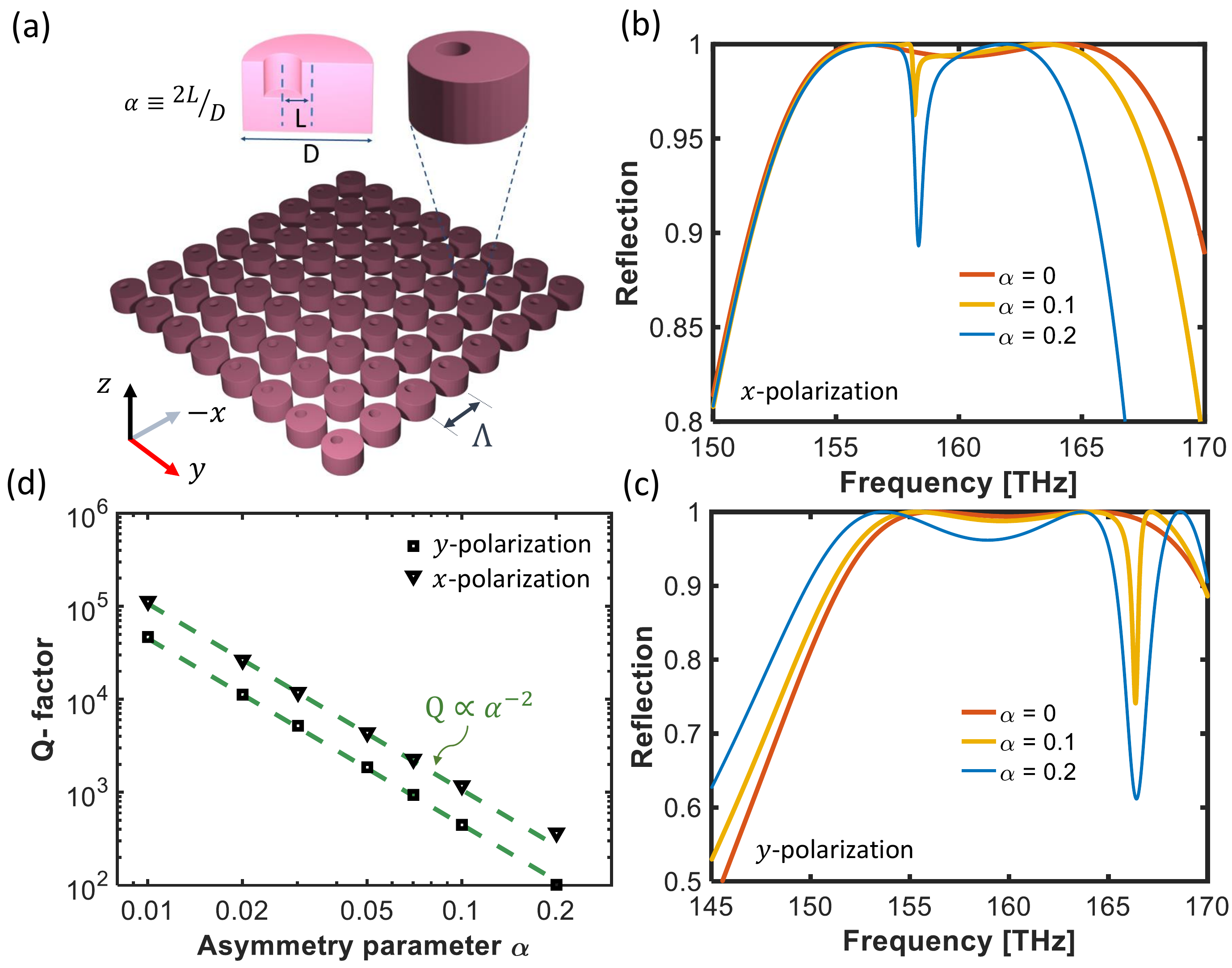}\caption{Linear response of the bianisotropic metasurface based on bound states in the continuum. (a) Schematic of the bianisotrapic metasurface (and its cross-sectional view) with a broken symmetry characterized by the asymmetry parameter $\alpha=2L/D$, where $ L$ is the shift distance between the center of the cylinder and the hole. (b-c) The reflection
spectrum with the $x$- and $y$-polarization of the illuminating light
for three different values of the asymmetry parameter $\alpha=0, 0.1, 0.2$. (d) The Q-factor as a function of the asymmetry parameter $\alpha$ for the $x$- and $y$-polarization
of the illuminating light obtained from the eigenmode analysis and  Eq.~(\ref{Qfactor}). The inverse quadratic dependence of the Q-factor on the asymmetry parameter $\,\rm Q\propto\alpha^{-2}$ shows that our proposed metasurface in (a) exhibits the symmetry-protected quasi-BICs~\cite{koshelev2018asymmetric}.}\label{fig2}
\end{figure*}

In this work, we demonstrate a highly \textit{asymmetric} generation of second harmonics assisted with quasi-BICs in nonlinear metasurfaces composed of bianisotropic AlGaAs meta-atoms~[see Fig.~\ref{fig1}(a) and Fig.~\ref{fig2}(a)]. We show that the asymmetric nonlinear response is originated from the electric and magnetic multipole moments with different strength when the nonlinear metasurface is illuminated from the forward and backward directions. Interestingly, we found a giant ratio of SHG
for the forward and backward excitation directions. Our results show that the ratio of SHG for the forward and backward excitation directions can be suitably controlled through the asymmetry parameter. In the following, we provide a solid methodological framework to explore the nonlinear features of the bianisotropic metasurfaces composed of $\,\rm Al_{0.18}Ga_{0.82}As$ meta-atoms with high relative permittivity~\cite{gehrsitz2000refractive} and large second-order nonlinear susceptibility ($\chi^{(2)}$), around 200 pm/V in the near-infrared spectral range~\cite{ohashi1993determination,gili2016monolithic}.


\section{Nonlinear bianisotropic metasurfaces}
Let us start with an AlGaAs bianisotropic metasurface composed of cylindrical meta-atom with a coaxial cylindrical hole with height $h$, see Fig.~\ref{fig1}(a). The metasurface is illuminated
by a linearly polarized light in $\pm z$ directions ($\mathbf{k}_{\rm FW/BW} = \pm k_0\mathbf{e}_z$, where + and - indicate forward and backward illuminations, respectively). Such a meta-atom design with magnetoelectric coupling shows an asymmetric linear response for the forward and backward illuminations~\cite{alaee2015all, odit2016experimental}. In particular, this asymmetry for \textit{lossless reciprocal} meta-atoms, manifests itself as different phases of the reflection coefficients
for the forward and backward illuminations~\cite{alaee2015all,odit2016experimental}, while the reflection amplitudes are identical for the forward and backward illuminations. The reflection coefficient in terms of the induced multipoles up to quadrupole orders is given by~\cite{terekhov2019multipole}
\begin{equation}
 r=\frac{i k_{0}}{\left(2\varepsilon_{0} E_{0}\Lambda^{2}\right)}\left[ p_{x}-\frac{ m_{y}}{c_{0}}+\frac{ik_{0}}{6}Q^{\rm e}_{xz}-\frac{i k_{0}}{6}\frac{ Q^{\rm m}_{yz}}{c_{0}}\right],\label{Eq_ref}
\end{equation}
where $\, k_{0}$, $\varepsilon_{0}$ and $\, E_{0}$ denote the wave vector in free space, permittivity of the vacuum and the amplitude of the incident electric field, respectively. $\varLambda$ is the periodicity in $x$ and $y$ direction. $p_x$ ($Q^{\rm e}_{xz}$), $m_y$ ($Q^{\rm m}_{yz}$) represent the
effective electric and magnetic dipole (quadrupole) moments, respectively~\cite{alaee2018electromagnetic}.

Figure~\ref{fig1}(b)-(c) show the reflection for the forward and backward illumination direction, respectively, obtained through the numerical simulation and semi-analytic method. The semi-analytic approach is based on  multipolar decomposition of the polarization current, using Eq.~(\ref{Eq_ref})~(see supporting information for details). Figure~\ref{fig1}(b)-(c) show a good agreement between the simulation and semi-analytic method. Although forward and backward reflections are identical, the multipolar contributions
are quite different for opposite illumination directions
(i.e., $\mathbf{k}_{\rm FW/BW}$). In particular, the magnetic dipole contribution becomes comparable to the electric dipole contribution
in the case of backward illumination. In fact, different field distributions in the meta-atom (or different polarization currents) give rise to various induced multipoles ~[see the insets of  Fig.~\ref{fig1}(b)-(c)]. 

To obtain the second-harmonic generation, we performed numerical simulations in two consecutive steps using COMSOL Multiphysics based on the finite element methods.  Firstly, Maxwell's equations are solved to calculate the field distributions and linear response of the bianisotropic structure.  Secondly, the nonlinear polarization current obtained through
the previous step is employed as a source of $2\omega$ in a scattering
problem in order to generate second harmonics. All three components of the nonlinear polarization vector for AlGaAs crystal, possessing a zinc-blende structure with symmetry $\overset{\_}{43}m$, can be
obtained as~\cite{boyd2020nonlinear}

\begin{equation}
\left[\begin{array}{c}
 P_{x}(2\omega)\\
 P_{y}(2\omega)\\
 P_{z}(2\omega)
\end{array}\right]=2\varepsilon_{0}\left[\begin{array}{cccccc}
0 & 0 & 0 & d & 0 & 0\\
0 & 0 & 0 & 0 & d & 0\\
0 & 0 & 0 & 0 & 0 & d
\end{array}\right]\left[\begin{array}{c}
 E^{2}(\omega)\\
 E^{2}(\omega)\\
 E^{2}(\omega)\\
 2E_{y}(\omega) E_{z}(\omega)\\
 2E_{x}(\omega) E_{z}(\omega)\\
 2E_{x}(\omega) E_{y}(\omega)
\end{array}\right],
\end{equation}

\noindent where $d_{14}=d_{25}=d_{36}=d$ are associated with the second-order
susceptibility tensor elements through the relation $2d(-2\omega;\omega,\omega)=\chi^{(2)}(-2\omega;\omega,\omega)$,
and the only nonvanishing elements are $\chi_{xyz}^{(2)}=\chi_{yzx}^{(2)}=\chi_{zxy}^{(2)}=\chi^{(2)}$.
 The second-harmonic
radiation generated by the bianisotropic metasurface is governed by
the diffraction orders because the periodicity of the lattice $\Lambda$
is larger than the radiated wavelength at $2\omega$. In the frequency domain the vectorial components of the nonlinear polarization current
can be calculated as $ J_{x}=-i(2\omega)\varepsilon_{0}\chi^{(2)} E_{y}(\omega) E_{z}(\omega), J_{y}=-i(2\omega)\varepsilon_{0}\chi^{(2)} E_{z}(\omega) E_{x}(\omega)$
and $ J_{z}=-i(2\omega)\varepsilon_{0}\chi^{(2)} E_{x}(\omega) E_{y}(\omega)$. 
\begin{figure*}
    \centering
    \includegraphics[width=0.98\textwidth]{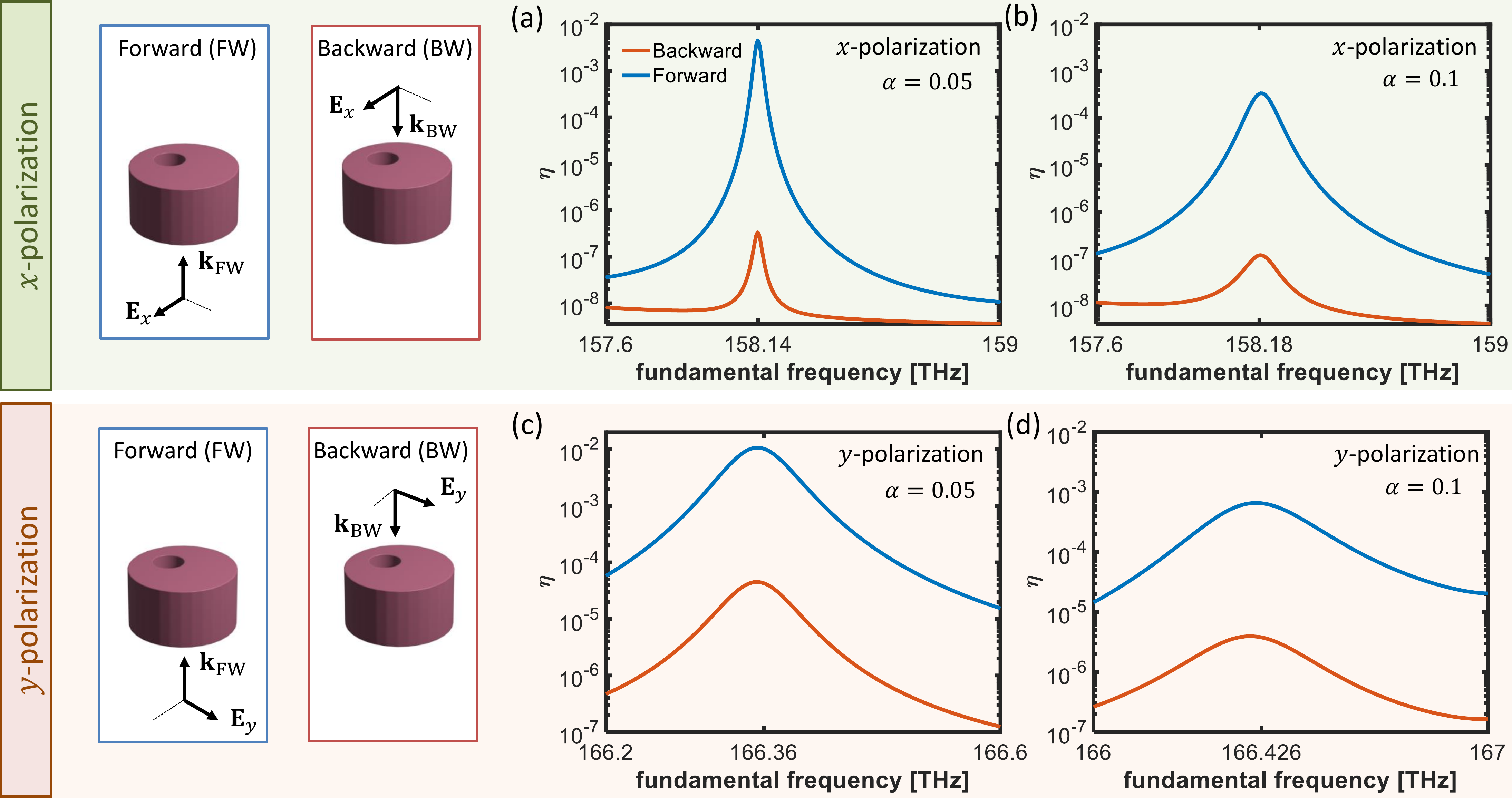}
\caption{Second harmonic (SH) conversion efficiencies of the bianisotropic metasurface based on bound states in the continuum. (a-b) The calculated SH conversion efficiencies for two different values of the asymmetry parameter, i.e., $\alpha=0.05$ and $0.1$ and $x-$polarization of illuminating light. The blue
and red curves denote forward and backward conversion efficiencies, respectively. The insets
show the $k$-vector and the polarization vector for different cases of illumination. (c-d) Same as (a-b) for $y-$polarization.}  \label{fig3}
\end{figure*}

The inset of Fig.~\ref{fig1}(b)-(c) show that  
the generated electric fields inside the bianisotropic
meta-atoms for the backward and forward illumination directions are different, leading to different nonlinear
polarization currents. Therefore, we expect to achieve different conversion efficiencies of
second harmonics $\,\rm \eta=P_{\rm SHG}/P_{\rm pump}$, where $P_{\rm SHG}$ and $P_{\rm pump}$ are the radiated power at $2\omega$ and the pump power at $\omega$, respectively. 
It can be seen in Fig.~\ref{fig1}(d) that the calculated conversion efficiency for the backward illumination is almost two times greater than that for the
forward illumination. Note that we assumed a pump intensity of $1\rm {MW}/\rm{cm^{2}}$.
  
\section{Nonlinear bianisotropic metasurfaces based on bound states in the continuum}
In the following, we propose a robust approach to enhance and control
the conversion efficiency of second harmonics for opposite illuminations
in bianisotropic metasurfaces. Optical BICs tightly connected with
sharp resonances are a powerful concept to fortify asymmetric features
of bianisotropic metasurfaces in the nonlinear regime. In the same
vein, we apply in-plane asymmetry to the bianisotropic meta-atom to
obtain quasi-BIC modes, and we calculate corresponding second-harmonic conversion
efficiencies. Let us consider a bianisotropic meta-atom with geometrical
dimension $d=270\,\rm{nm}$, $D=900\,\rm{nm}$, $h=360\,\rm{nm}$, $H=720\,\rm{nm}$,
and $\varLambda=1260\,\rm{nm}$, see Fig.~\ref{fig2}(a). The asymmetry parameter is defined as $\alpha=2 L/D$, where $ L$ is the shift distance between the center of the cylinder and the hole. 

Figure~\ref{fig2}(b) and (c) depict the reflection coefficients corresponding to three different values
of the asymmetry parameter and two polarizations of the illuminating
light. It can be seen from the figure
that the resonance linewidth increases as the asymmetry parameter
increases, and this facilitates the coupling channel of light to the
free space. Further, through the eigenmode analysis of the metasurface, the Q-factor can be obtained as 
\begin{equation}
Q=\frac{\,\rm Re(\omega_{r})}{2\,\rm Im(\omega_{r})}, \label{Qfactor}
\end{equation}
where $\omega_{r}$ is the complex eigenvalue. Figure~\ref{fig2}(d) illustrates
the Q-factor versus the asymmetry parameter for both $x$- and $y$-polarization, where the inverse quadratic dependence of the Q-factor on the asymmetry
parameter $\,\rm Q\propto\alpha^{-2}$ identifies the prominent resonances
as symmetry-protected quasi-BICs~\cite{koshelev2018asymmetric}. 

\begin{figure*}
\centering
\includegraphics[width=0.98\textwidth]{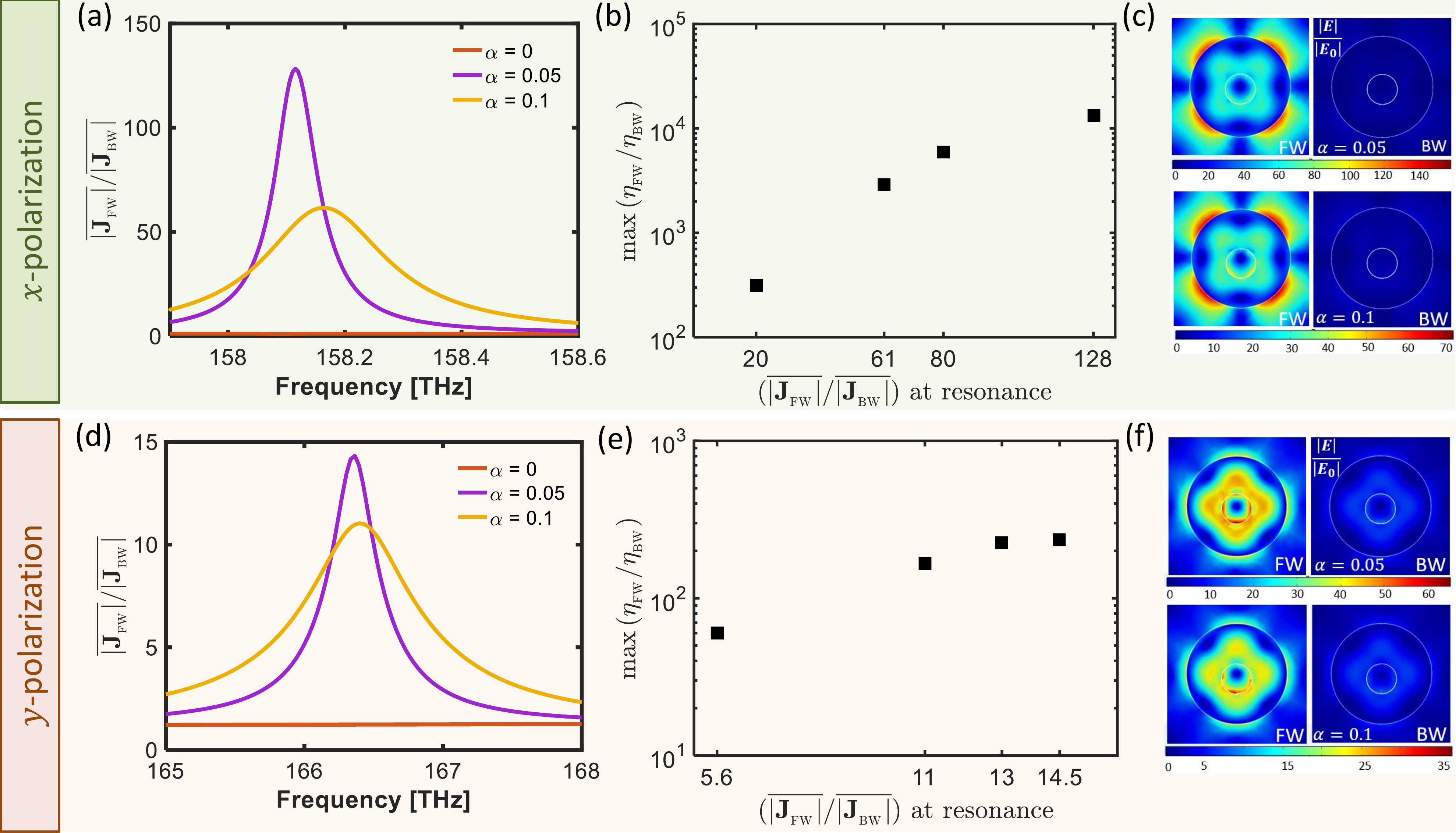}\caption{(a) the factor $\overline{|\mathbf{J}_{\rm FW}|}/\overline{|\mathbf{J}_{\rm BW}|}$ obtained from Eq.~(\ref{Eq_psi}) as a function of frequency for $x-$polarization and three different values of the asymmetry parameter, where the bar over current ($\mathbf{J_{\rm{FW/BW}}}$) indicates the average in the volume~[see Eq.~(\ref{Eq_psi})]. (b) Maximum ratio of $\left\{ {\,\rm \eta_{FW}/\eta_{BW}}\right\}$ as a function of factor $\overline{|\mathbf{J}_{\rm FW}|}/\overline{|\mathbf{J}_{\rm BW}|}$. Each $\overline{|\mathbf{J}_{\rm FW}|}/\overline{|\mathbf{J}_{\rm BW}|}$ (or each black square) is associated with a specific value of $\alpha$; the black squares are corresponding to $\alpha=0.2,0.1, 0.07, 0.05$, from left to right, respectively. (c) The normalized electric field in the middle of the meta-atom in $xy$-plane (shown in Fig.~\ref{fig2}(a)) for two different values of the asymmetry parameter, i.e., $\alpha=0.05$ and $0.1$ for $x-$ polarization. The normalized electric field distributions in the middle of the meta-atom are illustrated for forward and backward illumination direction at the resonance frequency. (d-f) Same as (a-c) for $y-$polarization. 
} \label{fig4}
\end{figure*}

To explore the asymmetric nonlinear response of our metasurface with quasi BICs, we calculated second-harmonic conversion efficiencies for opposite illumination directions, two different
values of the asymmetry parameter (i.e., $\alpha=0.05$ and $0.1$) and polarizations~[see Fig.~\ref{fig3}]. Our results demonstrate a high conversion efficiency and a giant ratio of imbalanced second harmonics for the opposite directions of illumination at the frequency of quasi-BIC modes. 

The second-harmonic conversion efficiency reduces as the asymmetry parameter increases
for both polarizations, occurring due to some increase in the resonance
linewidth~[e.g., compare Fig.~\ref{fig3}(a) and (b)]. We obtain a giant ratio of the forward to backward second-harmonic intensities ~[see the supporting information, $\,\rm \eta_{FW}/\eta_{BW}\approx 10^{4}$ for $x$-polarization]. This
ratio drops almost two orders of magnitude for the illuminating light
with $y$-polarization. Interestingly, being dual quasi-BICs with respect
to the polarization of the illuminating light adds one more degree
of freedom to the proposed design of the meta-atom for manipulation
and control of the ratio ${\rm \eta_{FW}/\eta_{BW}}$. 

To understand the underlying physical mechanism of 
the asymmetric second-harmonic generation in our bianisotropic metasurface, we explore the nonlinear polarization current as the source of second harmonics for opposite illumination directions. The relation between the average nonlinear polarization current at frequency $2\omega$ with the electric field components at the fundamental frequency can be expressed as 
\begin{widetext}
\begin{equation}
\overline{\left|\mathbf{J}_{j}\right|}=\frac{2\omega\varepsilon_{0}\chi^{\left(2\right)}}{V}\int_{V}\sqrt{\left|E_{x_{j}}\left(\omega\right)E_{y_{j}}\left(\omega\right)\right|^{2}+\left|E_{z_{j}}\left(\omega\right)E_{x_{j}}\left(\omega\right)\right|^{2}+\left|E_{y_{j}}\left(\omega\right)E_{z_{j}}\left(\omega\right)\right|^{2}}dV,\label{Eq_psi} 
\end{equation}
\end{widetext}
where $j=\left\{\rm BW,FW \right\} $ and the average in the volume of the meta-atom (V) is indicated by the bar above the current. Although the right hand side of Eq.(4) can be fully characterized in terms of electric field components obtained through the linear simulation, it provides an insight into the nonlinear optical response of the structure, through its relationship to the nonlinear polarization current components (i.e., $ J_{x},J_{y},J_{z}$). Figure~\ref{fig4}(a) and (d) show the factor
$\overline{|\mathbf{J}_{\rm FW}|}/\overline{|\mathbf{J}_{\rm BW}|}$ for three different values of the asymmetry parameter and $x-$ and $y-$ polarizations, respectively. It can be seen that the factor $\overline{|\mathbf{J}_{\rm FW}|}/\overline{|\mathbf{J}_{\rm BW}|}$ resonates at the frequency of the quasi-BIC modes for different polarizations. These resonances become weaker with the increase of the asymmetry parameter. Furthermore, through the calculations above for certain moderate values of the asymmetry parameter, i.e., $\alpha$, it can be realized that the ratio of second-harmonic power for the forward and backward directions rises when the factor $\overline{|\mathbf{J}_{\rm FW}|}/\overline{|\mathbf{J}_{\rm BW}|}$ increases (or equivalently when the asymmetry parameter decreases)~[see Fig.~\ref{fig4}(b) and Fig.~\ref{fig4}(e)]. Importantly, we observe the maximum induced nonlinear polarization
currents inside the meta-atoms at the quasi-BIC resonances coincide with the maximum asymmetry of second-harmonic generation. Basically, it indicates the presence of  great contrast between the field strength inside the meta-atoms for the forward and backward illuminations. To explore this, the normalised field distributions in the $xy$-plane (in the middle of the meta-atom) are shown in Fig.~\ref{fig4}(c) and Fig.~\ref{fig4}(f), with significantly different field enhancement factors for two opposite illumination directions. For the $x$-polarization ($y$-polarization), the field enhancement of 140 (60) and 70 (35) can be obtained with $\alpha=0.05$ and $0.1$, respectively, for the forward illuminating light, while these values drop significantly for the backward illuminating light~[compare Fig.~\ref{fig4}(c) and Fig.~\ref{fig4}(f)]. Fundamentally, it corroborates the collaborative role of bianisotropy and the quasi-BIC modes in the highly asymmetric generation of second harmonics in the proposed metasurface.

\section{Conclusions}
We theoretically proposed a giant asymmetric second harmonic generation through the combination of bianisotropy and quasi-BIC modes, originated
by broken in-plane symmetry in nonlinear metasurfaces. We analyzed our results based on the induced multipole moments and nonlinear polarization currents as the source of nonlinearity. We found that the sharp resonances of the quasi-BIC modes can be employed to enhance nonlinear polarization currents, coinciding with a very asymmetric field enhancement inside the meta-atoms caused by bianisotropy. Our approach to achieving asymmetric nonlinear response is \textit{not} restricted to second-harmonic generation and can be also extended to the third, higher-order harmonic generations, and parametric nonlinear processes~\cite{boyd2020nonlinear,liu2018all}. Our study has the potential to be used in nonlinear holography.

\section{Numerical simulations}

For simulating the reflection spectrum, the ratio of nonlinear polarization currents and the induced multipoles, we solve a full field problem in the frequency domain by COMSOL Multiphysics through using PML boundaries, periodic boundary conditions, and two exciting ports. Next, for the calculation of the conversion efficiencies, we solve a scattering problem by considering the nonlinear polarization current as the source of second harmonics at frequency
of $2\omega$. Finally, the nonlinear simulation is followed by calculating the second-harmonic power through the integration of the Poynting vector on two virtual planes placed at the top and the bottom of the unit cell, surrounding the meta-atom.

\textbf{Acknowledgments.—} R. A. acknowledges the support of the Alexander von Humboldt Foundation through the Feodor Lynen Return Research Fellowship. R.A. and R.W.B. acknowledge support through the Natural Sciences and Engineering Research Council of Canada, the Canada Research Chairs program, and the Canada First Research Excellence Fund. R.W.B. also thanks the US Army Research Office and the Office of Naval
Research for their support. K.D. acknowledges funding from Canada Research Chairs program, Natural Sciences and Engineering Research Council’s Discovery funding program [RGPIN-2020-03989], and the Canada First Research Excellence Fund. E. M. acknowledges Kaustubh Vias for his help with drawing the artistic figures.

\end{document}